\documentclass[11pt]{article}

\usepackage{amsfonts}
\usepackage{amssymb}
\usepackage{amsmath}
\usepackage{latexsym}
\usepackage{amsthm}
\usepackage{times}
\usepackage{fullpage}

\newcommand{\tinyspace}{\mspace{1mu}}

\newcommand{\norm}[1]{\left\lVert\tinyspace#1\tinyspace\right\rVert}
\newcommand{\abs}[1]{\left\lvert\tinyspace #1 \tinyspace\right\rvert}

\newcommand{\dnorm}[1]{\norm{#1}_{\diamond}}

\newcommand{\defeq}{\stackrel{\smash{\text{\tiny def}}}{=}}
\newcommand{\tr}{\operatorname{tr}}

\newcommand{\setft}[1]{\mathrm{#1}}
\newcommand{\lin}[1]{\setft{L}(\mathcal{#1})}

\newcommand{\trans}[1]{\setft{T}(\mathcal{#1})}

\newcommand{\ket}[1]{|\hspace{0.5pt}#1\hspace{0.5pt}\rangle}

\newcommand{\bra}[1]{\langle\hspace{0.5pt}#1\hspace{0.5pt}|}
\newcommand{\bracket}[2]{\langle\hspace{0.5pt}#1\hspace{0.5pt}|\hspace{0.5pt}
	#2\hspace{0.5pt}\rangle}
\newcommand{\opbracket}[3]{\langle\hspace{0.5pt}#1\hspace{0.5pt}|\hspace{0.5pt}
	#2\hspace{0.5pt}|\hspace{0.5pt}#3\hspace{0.5pt}\rangle}

\newenvironment{mylist}[1]{\begin{list}{}{
	\setlength{\leftmargin}{#1}
	\setlength{\rightmargin}{0mm}
	\setlength{\labelsep}{2mm}
	\setlength{\labelwidth}{8mm}
	\setlength{\itemsep}{0mm}}}
	{\end{list}}

\newcommand{\class}[1]{\textup{#1}}

\newcommand{\reg}[1]{\textsf{#1}}

\newtheorem{theorem}{Theorem}
\newtheorem{lemma}[theorem]{Lemma}

\newtheorem{cor}[theorem]{Corollary}
\theoremstyle{definition}
\newtheorem{defn}[theorem]{Definition}

\newtheorem{claim}[theorem]{Claim}

\begin{document}

%=============================================================================%

  \title{\Large
    {\bf Zero-knowledge against quantum attacks}\\
    (Preliminary version)
  }
  
  \author{
    John Watrous\\
    Department of Computer Science\\
    University of Calgary\\
    Calgary, Alberta, Canada
  }
  
  \date{November 3, 2005}
  
  \maketitle
  
  \begin{abstract}
    This paper proves that several interactive proof systems are zero-knowledge
    against quantum attacks.
    This includes a few well-known classical zero-knowledge proof
    systems as well as quantum interactive proof systems for the complexity
    class $\class{QSZK}_{\text{HV}}$, which comprises all problems having 
    ``honest verifier'' quantum statistical zero-knowledge proofs.
    It is also proved that zero-knowledge proofs for every language in 
    \class{NP} exist that are secure against quantum attacks, assuming the
    existence of quantum computationally concealing commitment schemes.
    Previously no non-trivial proof systems were known to be zero-knowledge
    against quantum attacks, except in restricted settings such as the 
    honest-verifier and common reference string models.
    This paper therefore establishes for the first time that true
    zero-knowledge is indeed possible in the presence of quantum information
    and computation.
  \end{abstract}
  
%=============================================================================%

\section{Introduction}

It is clearly to the benefit of honest users of a given cryptosystem that
security of the system is proved under as wide a range of malicious attacks as
possible.
At the same time it is desirable that honest users of the system are subjected
to as few resource requirements as possible.
The purpose of this paper is to investigate the security of zero-knowledge
proof systems against adversaries that use quantum computers to attack these
systems.
Although quantum interactive proof systems are considered in this paper,
our primary focus will be on the case of greatest practical interest, which
is the case where honest parties are not required to use quantum computers
to implement the proof systems.

The notion of zero-knowledge, first introduced in 1985 by Goldwasser, Micali
and Rackoff \cite{GoldwasserM+89}, is of central importance in theoretical
cryptography.
Informally speaking, an interactive proof system has the property of being
zero-knowledge if verifiers that interact with the honest prover of the
system learn nothing from the interaction beyond the validity of the statement
being proved.
At first consideration this notion may seem to be paradoxical, but indeed
several interesting computational problems that are not known to be
polynomial-time computable admit zero-knowledge interactive proof systems
in the classical setting.
Examples include the graph isomorphism \cite{GoldreichM+91} and quadratic
residuosity \cite{GoldwasserM+89} problems, certain lattice problems
\cite{GoldreichG00}, and the statistical difference \cite{SahaiV03}
and entropy difference \cite{GoldreichV99} problems that concern outputs of
boolean circuits with random inputs.
(The fact that the last three examples have interactive proof systems that are
zero-knowledge relies on a fundamental result of Goldreich, Sahai and Vadhan
\cite{GoldreichS+98} equating zero-knowledge with ``honest verifier'' 
zero-knowledge in certain settings.)
Under certain intractability assumptions, every language in $\class{NP}$
has a zero-knowledge interactive proof system \cite{GoldreichM+91}.
A related notion is that of an interactive argument, wherein
computational restrictions on the prover allow for zero-knowledge protocols
having somewhat different characteristics than protocols in the usual
interactive proof system setting \cite{BrassardC+88}.

Within the context of quantum information and computation it is natural to
consider the implications of the quantum model to the notion of zero-knowledge,
and indeed this has been a topic of investigation for several years.
Despite this fact, however, relatively little progress has been made---even
the first step of formulating a cryptographically reasonable, general
definition of quantum zero-knowledge and applying this definition to any
non-trivial computational problem was not taken previous to this paper.
The difficulty, which was apparently first discussed by van de Graaf
\cite{Graaf97}, has been that when the most natural quantum analogues of the
classical definitions of zero-knowledge were considered, the resulting
definitions appeared to be too strict to be applied to non-trivial proof
systems, including those systems already proved to be zero-knowledge in
the classical setting.
This has left open several possibilities, including the possibility that
any ``correct'' definition of quantum zero-knowledge would necessarily be
qualitatively different from the usual classical definitions, as well as the
possibility that zero-knowledge is simply impossible in a quantum world.

There are multiple classical variants of zero-knowledge that differ in the
specific way that the notion of ``learning nothing'' is formalized.
In each variant, it is viewed that a particular verifier learns nothing if
there exists a polynomial-time {\it simulator} whose output is
indistinguishable from the output of the verifier after interacting with
the prover on any positive instance of the problem.
The different variants concern the strength of this indistinguishability.
In particular, {\it perfect} and {\it statistical} zero-knowledge refer to
the situation where the simulator's output and the verifier's output are
indistinguishable in an information-theoretic sense and {\it computational}
zero-knowledge refers to the weaker restriction that the simulator's output
and the verifier's output cannot be distinguished by any computationally
efficient procedure.

It is straightforward to formulate fairly direct and natural quantum analogues
of the definitions of these classical variants of zero-knowledge.
Known proofs that specific proof systems are zero-knowledge with
respect to these classical definitions, however, do not translate directly
to the quantum setting.
The main obstacle when proving that a given proof system is zero-knowledge is
of course the construction of a simulator for every possible deviant
polynomial-time verifier.
Although there are different techniques for doing this, the most typical
method involves the simulator treating a given verifier as a 
{\it black box}: the simulator randomly produces transcripts, or parts of
transcripts, of possible interactions between a prover and verifier, and feeds
parts of these transcripts to the given verifier.
If the verifier produces a message that is not consistent with the
other parts of the transcript that were produced, the simulator ``rewinds'',
meaning that it backs up and tries again to randomly generate parts of the
transcript.
By storing intermediate results, and repeating different parts of this
process until the given verifier's output is consistent with a randomly
generated transcript, the simulation is eventually successful.
The reason why this technique cannot generally be applied directly to quantum
verifiers is based on the facts that (i) quantum information cannot be
copied, and (ii) measurements are irreversible processes---their effects
cannot in general be undone.
If a simulator runs a given verifier as a black box and the simulation is
unsuccessful, it is not clear how to rewind the process and try
again; intermediate states of the system cannot be copied, and running
the verifier may have involved an irreversible measurement.
More significantly, the determination of whether the simulation was successful
will itself represent an irreversible measurement in general.
Other methods of constructing simulators for quantum verifiers have also
not been successful in the general setting.
Further discussions of this issue can be found in \cite{Graaf97} and
\cite{DamgardF+04}.

There are several ``weaker'' notions of zero-knowledge that are of interest
and have been studied, both in the quantum and classical cases. 
Of particular interest with respect to previous work on quantum zero-knowledge
is the {\it common reference string} model, wherein it is assumed that an
honest third party samples a string from some specified distribution and
provides both the prover and verifier with this string at the start of the
interaction.
Damg{\aa}rd, Fehr, and Salvail \cite{DamgardF+04} proved several interesting
results concerning quantum zero-knowledge protocols in this context.
Their results are centered on what they call the {\it no quantum rewinding
paradigm}, where the central issue concerning simulator constructions discussed
above is partially circumvented by making use of common reference
strings as well as certain unproved quantum complexity-theoretic assumptions.
Their results are also mostly concerned with interactive arguments, which
require computational restrictions on the prover to establish soundness.
Another weaker notion of zero-knowledge is ``honest verifier'' zero-knowledge,
which only requires a simulator that outputs the verifier's view of the
interaction between the honest parties $V$ and $P$.
A quantum variant of honest verifier statistical zero knowledge was considered
in \cite{Watrous02}, wherein it was proved that the resulting complexity
class shares many of the basic properties of its classical counterpart
\cite{SahaiV03}.
A non-interactive variant of this notion was studied by
Kobayashi~\cite{Kobayashi03}.
The problematic issue regarding simulator constructions does not occur in
honest verifier settings.

The present paper essentially resolves the main difficulties previously
associated with quantum analogues of zero-knowledge.
This is done by establishing that the most natural quantum analogues of the
classical definitions of zero-knowledge indeed can be applied to a large class
of proof systems.
This includes several well-known classical proof systems as well as quantum
proof systems for many problems, in particular the class of all problems
admitting quantum proof systems that are statistical zero-knowledge against
honest verifiers.
We therefore prove unconditionally that zero-knowledge indeed is possible in
the presence of quantum information and computation, and moreover that the
notion of quantum zero-knowledge is correctly captured by the most natural
and direct quantum analogues of the classical definitions.
The basic technique we use in the paper is algorithmic in nature: we show how
to construct efficient quantum simulators for arbitrary quantum polynomial-time
deviant verifiers for several proof systems.
The proof that these simulators operate correctly involves a simple spectral
property of measurement operators resulting from the most straightforward
(but not always successful) simulator constructions, combined with a fact
that previously was used to reduce errors in $\class{QMA}$ proof systems
without increasing witness sizes \cite{MarriottW05}.

The remainder of this paper is organized as follows.
Section~\ref{sec:preliminaries} discusses definitions of zero-knowledge,
including standard classical definitions and quantum analogues
of these definitions.
Section~\ref{sec:graph-isomorphism} focuses on the well-known zero-knowledge
graph isomorphism protocol of Goldreich, Micali, and Wigderson
\cite{GoldreichM+91}, proving that this protocol is zero-knowledge against
quantum attacks.
It is intended that this proof illustrates, in a simple and familiar
setting, a more general method that can be applied to several other
protocols.
Some other protocols that can be proved zero-knowledge using this method
are discussed in Section~\ref{sec:other}.
The paper concludes with Section~\ref{sec:conclusion}, which 
mentions some possible directions for
future work.

%=============================================================================%

\section{Definitions of Zero-Knowledge} \label{sec:preliminaries}

This paper assumes the reader is familiar with the notions of interactive
proof systems, zero-knowledge, and quantum computation.
Further information on interactive proof systems and zero knowledge can
be found, for instance, in \cite{Goldreich01,Goldreich02}.
Standard references for quantum computation and information include
\cite{NielsenC00, KitaevS+02}.
Quantum computational variants of interactive proof systems were studied
in \cite{Watrous03-pspace, KitaevW00}.

In this paper, interactive proof systems will be specified by pairs $(V,P)$
representing honest verifier and honest prover strategies.
The soundness property of such an interactive proof system concerns
interactions between pairs $(V,P')$ and the zero-knowledge property concerns
interactions between pairs $(V',P)$, where $P'$ and $V'$ deviate adversarily
from $P$ and $V$, respectively.
It may be the case that a given pair of interacting strategies is such that
both are classical, both are quantum, or one is classical and the other is
quantum.
When either or both of the strategies is classical, all communication between
them is (naturally) assumed to be classical---only two quantum strategies
are permitted to transmit quantum information to one another.
It will always be assumed that verifier strategies are represented by
polynomial-time (quantum or classical) computations.
Depending on the setting of interest, the honest prover strategy $P$ may either
be computationally unrestricted or may be represented by a polynomial-time
(quantum or classical) computation augmented by specific information about the
input string, such as a witness for an \class{NP} problem.
Deviant prover strategies $P'$ will always be assumed to be computationally
unrestricted.
(A proof system $(V,P)$ for which the soundness property requires a 
computational assumption on $P'$ is called an {\it interactive argument}
\cite{BrassardC+88}.
Although the results of the present paper are applicable to interactive
arguments, none are specific to arguments, so for simplicity they are not
discussed further.)

For a given promise problem $A = (A_{\mathrm{yes}},A_{\mathrm{no}})$,
we say that a pair $(V,P)$ is an interactive proof system for $A$ having
completeness error $\varepsilon_c$ and soundness error $\varepsilon_s$ if
(i) for every input $x\in A_{\mathrm{yes}}$, the interaction between $P$ and
$V$ causes $V$ to accept with probability at least $1 - \varepsilon_c$, and
(ii)
for every input $x\in A_{\mathrm{no}}$ and every prover strategy $P'$, the
interaction between $P'$ and $V$ causes $V$ to accept with probability at
most $\varepsilon_s$.
It may be the case that $\varepsilon_c$ and $\varepsilon_s$ are constant or are
functions of the length of the input string $x$.
When they are functions, it is assumed that they can be computed
deterministically in polynomial time.
It is generally desired that $\varepsilon_c$ and $\varepsilon_s$ be
exponentially small.
As sequential repetition followed by majority vote, or unanimous vote in case
$\varepsilon_c = 0$, reduces these errors exponentially quickly, it is usually
sufficient that $1 - \varepsilon_c - \varepsilon_s$ is lower-bounded by the
reciprocal of a polynomial.
(The same may be said of parallel repetition, but the zero-knowledge property
to be discussed shortly will generally be lost in this case.)

There are different notions of what it means for an interactive proof
system $(V,P)$ for a promise problem $A$ to be zero-knowledge.
Let us first discuss the completely classical case, meaning that only classical
strategies are considered for the honest verifier $V$ and any deviant verifiers
$V'$.
An arbitrary verifier $V'$ takes two strings as input---a string
$x$ representing the common input to both the verifier and prover, as well as
a string $w$ called an {\it auxiliary input}, which is not known to the prover
and which may influence the verifier's behavior during the interaction.
Based on the interaction with $P$, the verifier $V'$ produces a string
as output.
Let $n,m:\{0,1\}^{\ast}\rightarrow\mathbb{N}$ be polynomially-bounded
functions representing the length of the auxiliary input string and output
string: assuming the common input string is $x$, the auxiliary input is a
string of length $n(x)$ and the output is a string of length $m(x)$.
Because there may be randomness used by either or both of the strategies
$P$ and $V'$, the verifier's output will in general be random.
The random variable representing the verifier's output will be written
$(V'(w),P)(x)$.
For the honest verifier $V$, we may view that $n = 0$ and $m = 1$, because
there is no auxiliary input and the output is a single bit that indicates
whether the verifier accepts or rejects.

By a (classical) simulator we mean a polynomial-time randomized algorithm $S$
that takes strings $w$ and $x$, with $\abs{w} = n(x)$, as input and
produces some output string of length $m(x)$.
Such a simulator's output is a random variable denoted $S(w,x)$.
Now, for a given promise problem $A$, we say that a proof system $(V,P)$
for $A$ is zero-knowledge if, for every verifier $V'$ there exists a 
simulator $S$ such that $(V'(w),P)(x)$ and $S(w,x)$ are indistinguishable
for every choice of strings $x\in A_{\mathrm{yes}}$ and $w\in\{0,1\}^{n(x)}$.
The specific formalization of the word ``indistinguishable'' gives rise to
different variants of zero-knowledge.
{\it Statistical} zero-knowledge refers to the situation in which $(V(w),P)(x)$
and $S(w,x)$ have negligible statistical difference, and {\it computational}
zero-knowledge refers to the situation in which no boolean circuit with size
polynomial in $\abs{x}$ can distinguish $(V'(w),P)(x)$ and $S(w,x)$ with
a non-negligible advantage over randomly guessing.
({\it Perfect} zero-knowledge is slightly stronger than statistical
zero-knowledge in that it essentially requires a zero-error simulation: the
simulator may report failure with small probability, and conditioned on the
simulator not reporting failure the outputs $S(w,x)$ and $(V'(w),P)(x)$ are
identically distributed.)

Two points concerning the above definitions should be mentioned.
The first point concerns the auxiliary input, which actually was not included
in the definitions given in the very first papers on zero-knowledge (but
which already appeared in the 1989 journal version of
\cite{GoldwasserM+89}).
The inclusion of an auxiliary input in the definition is needed to prove
that zero-knowledge proof systems are closed under sequential composition.
Perhaps more important is that the inclusion of auxiliary inputs in the
definition captures the notion that a given zero-knowledge proof system
cannot be used to {\it increase} knowledge.
The second point concerns the order of quantification between $V'$ and $S$.
Specifically, the definition states that a zero-knowledge proof system is
one such that for all $V'$ there exists a simulator $S$ that satisfies the
requisite properties.
There is a good argument to be made for reversing these quantifiers by
requiring that for a given proof system $(V,P)$ there should exist a single
simulator $S$ that interfaces in some uniform way with any given
$V'$ to produce an output that is indistinguishable from that verifier's
output.
Typical simulator constructions, as well as the ones that will be discussed
in this paper in the quantum setting, do indeed satisfy this stronger
requirement.

Next let us discuss the case where a given deviant verifier strategy $V'$
may be quantum.
This includes the possibility that $V$ is classical or quantum, and likewise
for $P$.
Similar to the completely classical case, a given strategy $V'$ will take, in
addition to the input string $x$, an auxiliary input, and produce some output.
The most general situation allowed by quantum information theory is that both
the auxiliary input and the output are quantum states.
Moreover, it may be the case that the auxiliary input state qubits are
entangled with some other qubits that {\it are not} accessible to the verifier
or simulator, but {\it are} available to any procedure that attempts to
distinguish between the verifier and simulator outputs.
It is intended that this is a strong assumption, but it can easily be
argued that no sensible definition would forbid this possibility;
one can imagine natural situations in which potential attacks
could be based on entangled states in the sense described.

Similar to the classical case, it will be assumed that for every verifier
strategy $V'$ there exist polynomially bounded functions $n$ and $m$ that 
specify the number of auxiliary input qubits and output qubits of $V'$.
The interaction of $V'$ with $P$ on input $x$ is a physical process, and
therefore induces some {\it admissible} mapping $\Phi_x$ from $n(x)$ qubits
to $m(x)$ qubits.
This means that
$\Phi_x: \lin{\mathcal{W}}\rightarrow \lin{\mathcal{Z}}$
is a completely positive and trace preserving linear map, where
$\mathcal{W}$ and $\mathcal{Z}$ are Hilbert spaces corresponding to the
$n(x)$ auxiliary input qubits and the $m(x)$ output qubits, and
$\lin{\mathcal{W}}$ and $\lin{\mathcal{Z}}$ denote the spaces of linear
operators (including the density operators) acting on $\mathcal{W}$ and
$\mathcal{Z}$, respectively.
Likewise, a simulator $S$ given by some polynomial-time quantum computation
that takes as input the string $x$ along with $n(x)$ auxiliary input qubits
and outputs $m(x)$ qubits will give rise to some
admissible mapping $\Psi_x :\lin{\mathcal{W}}\rightarrow \lin{\mathcal{Z}}$.

We may now define variants of zero-knowledge based on different notions of
indistinguishability of these mappings $\Phi_x$ and $\Psi_x$.
The correct quantum analogue of statistical zero-knowledge requires that
$\dnorm{\Phi_x - \Psi_x}$ is negligible, where $\dnorm{\,\cdot\,}$ denotes
Kitaev's ``diamond'' norm \cite{Kitaev97,KitaevS+02,AharonovK+98}.
Informally this implies that no physical process can distinguish
$\Phi_x$ and $\Psi_x$ given a single ``black-box'' access to one of the
two mappings, including the possibility that the mapping is applied to just
one part of a larger, possibly entangled state.
Under the assumption that $\dnorm{\Phi_x - \Psi_x}$ is negligible,
it can be argued that no polynomial number of black-box accesses to
$\Phi_x$ or $\Psi_x$ would suffice to distinguish the two with non-negligible
probability.
Computational zero-knowledge is formulated similarly, except that the
distinguishing procedure must be specified by a polynomial-size quantum
circuit.
Because we only discuss quantum computational zero-knowledge in 
Section~\ref{sec:other}, a more precise definition will be postponed until
then.

%=============================================================================%

\section{The Goldreich-Micali-Wigderson Graph Isomorphism Proof System}
\label{sec:graph-isomorphism}

The Goldreich-Micali-Wigderson graph isomorphism protocol is a well-known
example of a proof system that is perfect zero-knowledge against classical
polynomial-time verifiers.
In this section it is proved that this protocol is in fact zero-knowledge
against polynomial-time quantum verifiers.
The purpose of focusing on this particular protocol is that it provides a
simple example where nevertheless the difficulties associated
with constructing simulators for quantum verifiers are present.
The proof that this proof system remains zero-knowledge against quantum
attacks also illustrates clearly the basic method being proposed in this
paper.
The method can be extended to several other protocols as described in the
next section.

The protocol is described in Figure~\ref{fig:protocol}.
It has perfect completeness and soundness error 1/2;
if $G_0\cong G_1$, then $V$ accepts with certainty, while if $G_0\not\cong G_1$
then no prover $P'$ can convince $V$ to accept with probability greater than
1/2.
The proof system has the property that if the prover $P$ has knowledge of
an isomorphism $\sigma:G_1\mapsto G_0$, then it may be taken to run in
polynomial time.
For an arbitrary choice of $\sigma$, the proof system $(V,P)$ is perfect
zero-knowledge with respect to any classical polynomial-time verifier $V'$.
Sequential repetition followed by a unanimous vote can be used to decrease the
soundness error to an exponentially small quantity while preserving the perfect
completeness and classical zero-knowledge properties.

\begin{figure}[ht]
\noindent\hrulefill
\vspace{2mm}

{\bf\large\centerline{Zero-Knowledge Protocol for Graph Isomorphism}}

\begin{trivlist}

\item Assume the input is a pair $(G_0,G_1)$ of simple, undirected graphs each
  having vertex set $\{1,\ldots,n\}$.

\item {\bf Prover's step 1:}
  Let $\sigma\in S_n$ be a permutation satisfying $\sigma(G_1) = G_0$ if
  $G_0 \cong G_1$, and let $\sigma$ be the identity permutation otherwise.
  Choose $\pi\in S_n$ uniformly at random and send $H = \pi(G_0)$ to the
  verifier.

\item {\bf Verifier's step 1:}
  Choose $a\in\{0,1\}$ uniformly at random and send $a$ to the prover.

\item {\bf Prover's step 2:}
  Let $\tau = \pi \sigma^a$ and send $\tau$ to the verifier.

\item {\bf Verifier's step 2:}
  Accept if $\tau(G_a) = H$, reject otherwise.
\end{trivlist}
\caption{The Goldreich, Micali, and Wigderson~\cite{GoldreichM+91}
  zero-knowledge graph isomorphism protocol.}
\label{fig:protocol}
\noindent\hrulefill
\end{figure}

We wish to show that this protocol is zero-knowledge with respect to
polynomial-time {\it quantum} verifiers.
It will be sufficient to consider a restricted type of verifier as follows:

\begin{mylist}{\parindent}
\item[$\bullet$]
In addition to $(G_0,G_1)$, the verifier takes a quantum register $\reg{W}$ as
input, representing the auxiliary quantum input.
The verifier will use two additional quantum registers that function as work
space: $\reg{V}$, which is an arbitrary  (polynomial-size) register, and
$\reg{A}$, which is a single qubit register.
The registers $\reg{V}$ and $\reg{A}$ are initialized to their all-zero states
before the protocol begins.

\item[$\bullet$]
In the first message, the prover $P$ sends a graph $H$ on $n$ vertices.
For each graph $H$ there corresponds a unitary operator $V_H$ that the verifier
applies to the registers $(\reg{W},\reg{V},\reg{A})$.
After applying the appropriate transformation $V_H$, the verifier measures
the register $\reg{A}$ with respect to the standard basis, and sends the
resulting bit $a$ to the prover.

\item[$\bullet$]
The prover responds with some permutation $\tau\in S_n$.
Because we are concerned only with the zero-knowledge properties of the
system, we assume the verifier does not make a decision to accept or reject,
but simply outputs the registers $(\reg{W},\reg{V},\reg{A})$, along with
the classical messages $H$ and $\tau$ sent by the prover during the protocol.

\end{mylist}

\noindent
Note that a verifier of this form is completely determined by the
collection $\{V_H\}$.

An arbitrary verifier can be modeled as a verifier of this restricted form
followed by some polynomial-time post-processing of this verifier's output.
The same post-processing can be applied to the output of the simulator that
will be constructed for the given restricted verifier.
We also note that it can be shown that a sequential repetition of the proof
system will also result in a zero-knowledge proof system against quantum
verifiers, based on the fact that the auxiliary input may be an arbitrary
quantum state.

%-----------------------------------------------------------------------------%

\subsection*{The mapping induced by the actual interaction}

Let us begin by considering the admissible transformation induced by an
interaction of a verifier of the above type with the prover $P$ in the case
that $G_0\cong G_1$.
Although the messages sent from the prover to the verifier are classical
messages, it will simplify matters to view them as being stored in quantum
registers denoted $\reg{Y}$ and $\reg{Z}$, respectively.
(Later, when we consider simulations of the interaction, we will need quantum
registers to store these messages anyway, and it is helpful to have the
registers used in the actual protocol and in the simulation share the same
names.)
With each register we associate a Hilbert space, and use the same letter in
different fonts for matching registers and spaces---for example,
$\mathcal{W}$ is the space associated with register $\reg{W}$,
$\mathcal{V}$ is the space associated with register $\reg{V}$, and so on.
Let $\ket{0_{\mathcal{V}\otimes\mathcal{A}}}\in\mathcal{V}\otimes\mathcal{A}$
denote the initial all-zero state of the registers $(\reg{V},\reg{A})$.
Let us also write $\mathcal{G}_n$ to denote the set of all simple, undirected
graphs having vertex set $\{1,\ldots,n\}$.

For each $H\in\mathcal{G}_n$ and each $a\in\{0,1\}$, define a linear mapping
$M_{H,a}\in\lin{\mathcal{W},\mathcal{W}\otimes\mathcal{V}}$ as
\[
M_{H,a} = \left( I_{\mathcal{W}\otimes\mathcal{V}} \otimes \bra{a}\right)
V_H (I_{\mathcal{W}} \otimes \ket{0_{\mathcal{V}\otimes\mathcal{A}}}).
\]
If the initial state of the register $\reg{W}$ is a pure state
$\ket{\psi}\in\mathcal{W}$, then the state of the registers
$(\reg{W},\reg{V},\reg{A})$ after the verifier applies $V_H$ is
$(M_{H,0} \ket{\psi})\ket{0} + (M_{H,1} \ket{\psi})\ket{1}$,
and therefore the state of the registers 
$(\reg{W},\reg{V},\reg{A})$ after the verifier applies $V_H$ and measures
$\reg{A}$ in the standard basis is
\[
\sum_{a \in\{0,1\}}
M_{H,a} \ket{\psi}\bra{\psi} M_{H,a}^{\ast} \otimes \ket{a}\bra{a}.
\]

The admissible map that results from the interaction is now easily
described by incorporating the description of $P$.
It is described by $\Phi \in \trans{\mathcal{W},
\mathcal{W}\otimes\mathcal{V}\otimes\mathcal{A}\otimes\mathcal{Y}
\otimes\mathcal{Z}}$ given by
\begin{equation} \label{eq:actual-interaction}
\Phi(X) = \frac{1}{n!} \sum_{\pi\in S_n} \sum_{a\in\{0,1\}}
M_{H,a} X M_{H,a}^{\ast} \otimes \ket{a}\bra{a}
\otimes \ket{\pi(G_0)}\bra{\pi(G_0)} \otimes 
\ket{\pi \sigma^a}\bra{\pi\sigma^a}
\end{equation}
for all $X\in\lin{\mathcal{W}}$.

%-----------------------------------------------------------------------------%

\subsection*{Description of the simulator}

A classical simulation for a classical verifier $V'$ in the above protocol may
be obtained as follows.
The simulator randomly choose a permutation $\pi$ and a bit $b$, and
feeds $\pi(G_b)$ to $V'$.
This verifier chooses a bit $a$ for its message back to the prover.
If $a = b$, the simulator can easily complete the simulation,
otherwise it ``rewinds'' and tries a new choice of $\pi$ and $b$.
With very high probability, the simulator will succeed after no more
than a polynomial number of steps.

Our procedure for simulating the verifier described by a collection
$\{V_H\,:\,H\in\mathcal{G}_n\}$ will require two registers $\reg{B}$ and
$\reg{R}$ in addition to $\reg{W}$, $\reg{V}$, $\reg{A}$, $\reg{Y}$,
and~$\reg{Z}$.
The register $\reg{R}$ may be viewed as a quantum register whose basis states
correspond to the possible random choices that a typical classical simulator
would use.
In the present case this means a random permutation together with a random bit.
The register $\reg{B}$ will represent the simulator's ``guess'' for the
verifier's message.
For convenience, let us define
$\mathcal{X} = 
\mathcal{V}\otimes\mathcal{A}\otimes\mathcal{Y}\otimes\mathcal{B}\otimes
\mathcal{Z}\otimes \mathcal{R}$,
which is the Hilbert space corresponding to all registers aside from $\reg{W}$,
and let $\ket{0_{\mathcal{X}}}$ denote the all-zero state of these registers.

The procedure, which is described in Figure~\ref{fig:simulator},
will involve a composition of a few operations that we now describe.
First, let $T$ be any unitary operator acting on registers
$(\reg{Y},\reg{B},\reg{Z},\reg{R})$ that maps the initial all-zero state
of these four registers to the state
\[
\frac{1}{\sqrt{2 n!}} \sum_{b\in\{0,1\}}\sum_{\pi\in S_n}
\ket{\pi(G_b)}\ket{b}\ket{\pi}\ket{\pi,b}.
\]
If the space corresponding to register $\reg{R}$ is traced out, the state
of registers $(\reg{Y},\reg{B},\reg{Z})$ corresponds to a classical
probability distribution over triples $(\pi(G_b),b,\pi)$ for $b$ and $\pi$
chosen uniformly.
Next, define a unitary operator $V$ acting on registers
$(\reg{W},\reg{V},\reg{A},\reg{Y})$ that effectively uses $\reg{Y}$ as
a control register, applying $V_H$ to registers $(\reg{W},\reg{V},\reg{A})$
for each possible graph $H\in\mathcal{G}_n$ representing a standard basis
state of $\reg{Y}$.
In other words,
$V = \sum_{H\in\mathcal{G}_n} V_H\otimes \ket{H}\bra{H}$.
The operators $T$ and $V$ are tensored with the identity on the remaining
spaces when we wish to view them both as operators on $\mathcal{W}\otimes
\mathcal{X}$.

Finally, we need to define a binary-valued projective measurement on the
above registers whose outcome is the exclusive-OR of $\reg{A}$ and $\reg{B}$
(with respect to the standard basis).
We will let $\Pi_0$ and $\Pi_1$ denote the projection operators corresponding
to this measurement.
Viewed as projections on $\mathcal{A}\otimes\mathcal{B}$, these projections
can be written $\Pi_0 = \ket{00}\bra{00} + \ket{11}\bra{11}$ and
$\Pi_1 = \ket{01}\bra{01} + \ket{10}\bra{10}$.
It will be more convenient, however, to view $\Pi_0$ and $\Pi_1$ as
projections on the entire space $\mathcal{W}\otimes\mathcal{X}$, so the
above projections should be tensored with the identity operator on
$\mathcal{W}\otimes\mathcal{V}\otimes\mathcal{Y}\otimes\mathcal{Z}\otimes
\mathcal{R}$.

\begin{figure}[ht]

\noindent\hrulefill
\vspace{3mm}

{\bf\large\centerline{Simulation Procedure}}
\begin{tabbing}
xxx\=xxx\=xxx\=xxxxx\=\kill
{\it Input and initial conditions:}\+\\[1mm]
The auxiliary input is register $\reg{W}$.\\
The registers $\reg{V}$, $\reg{A}$, $\reg{Y}$, $\reg{B}$, $\reg{Z}$, and
$\reg{R}$ are initialized to the state $\ket{0_{\mathcal{X}}}$.\\[1mm]
\-\kill
{\it Main procedure:}\+\\[1mm]
Perform the unitary transformation $T$ followed by the unitary 
transformation $V$.
\\[1mm]
Perform the measurement described by $\{\Pi_0,\Pi_1\}$.\\[1mm]
If the outcome of the measurement is 1:\\[1mm]
\+\kill
Perform the unitary transformation $V^{\ast}$ followed by $T^{\ast}$.\\[1mm]
Perform a phase flip in case any of the qubits in any of the registers
aside from $\reg{W}$ is not set to 0.\\
In other words, perform the unitary transformation
$I_{\mathcal{W}}\otimes (2 \ket{0_{\mathcal{X}}}
\bra{0_{\mathcal{X}}} - I_{\mathcal{X}})$.\\[1mm]
Perform the unitary transformation $T$ followed by the unitary 
transformation $V$.\\[1mm]
\-\kill
Halt and output registers $(\reg{W},\reg{V},\reg{A},\reg{Y},\reg{Z})$.
(Registers $\reg{B}$ and $\reg{R}$ are traced out.)
\end{tabbing}\vspace{-3mm}
\caption{The simulation procedure for the graph isomorphism protocol.}
\label{fig:simulator}
\noindent\hrulefill
\end{figure}

%-----------------------------------------------------------------------------%

\subsection*{Analysis of the simulator}

Now we will consider the mapping induced by the simulation procedure
described in Figure~\ref{fig:simulator}.
We are only concerned with its behavior in the case that $G_0\cong G_1$,
so this assumption is made hereafter.

We will first analyze the behavior of the simulator in the case that
the register $\reg{W}$ is in a pure state $\ket{\psi}\in\mathcal{W}$.
The analysis of the general case where $\reg{W}$ is in a mixed state, and
is possibly entangled with other registers not accessible to the simulator,
will follow easily from the pure state case.
The remaining registers $(\reg{V},\reg{A},\reg{Y},\reg{B},\reg{Z},\reg{R})$
begin the simulation in state $\ket{0_{\mathcal{X}}}$.
The initial state of all of the registers together is therefore
$\ket{\gamma_0} \defeq \ket{\psi}\ket{0_{\mathcal{X}}}$.
The simulator first performs the unitary transformation $T$ followed by $V$,
which transforms the state of the system to
\[
VT \ket{\gamma_0} = 
\frac{1}{\sqrt{2 n!}}
\sum_{a,b\in\{0,1\}} \sum_{\pi\in S_n}
(M_{\pi(G_b),a} \ket{\psi})
\ket{a}
\ket{\pi(G_b)}
\ket{b}
\ket{\pi}
\ket{\pi,b}.
\]
Next, the measurement $\{\Pi_0,\Pi_1\}$ is performed.
There are two possible measurement outcomes that will be considered
separately.

The easier case is that the measurement outcome is 0, in which case the
simulator does nothing more than to output
$(\reg{W},\reg{V},\reg{A},\reg{Y},\reg{Z})$, discarding the remaining two
registers $\reg{B}$ and $\reg{R}$.
Using the fact that $\pi(G_0) = \pi\sigma^b(G_b)$, we see that
\begin{align*}
\tr_{\mathcal{B}\otimes\mathcal{R}}\left(
\Pi_0 V T \ket{\gamma_0}\bra{\gamma_0} T^{\ast} V^{\ast} \Pi_0\right)
\hspace{-42mm}\\
& =
\frac{1}{2 n!}
\sum_{a\in \{0,1\}} \sum_{\tau\in S_n}
M_{\tau(G_0),a} \ket{\psi}\bra{\psi} M_{\tau(G_0),a}^{\ast}
\otimes \ket{a}\bra{a}
\otimes
\ket{\tau(G_0)}\bra{\tau(G_0)}
\otimes
\ket{\tau \sigma^a}\bra{\tau \sigma^a}\\
& = \frac{1}{2}\Phi(\ket{\psi}\bra{\psi}),
\end{align*}
where $\Phi$ is the admissible map of Eq.~\ref{eq:actual-interaction} that
corresponds to an actual interaction of $V$ with $P$.
It therefore holds that the simulator is correct conditioned on the measurement
outcome being 0.
Let us also note that the probability associated with outcome 0 is
the trace of this operator, which is $1/2$.
Of course it is intuitive that this probability should be 1/2, and that the
simulator is correct in this case---the more difficult case is when the
measurement result is 1.

In order to understand the behavior of the simulator in case the measurement
result is 1, it will be helpful to define two additional projections:
$\Delta_0 = I_{\mathcal{W}}\otimes \ket{0_{\mathcal{X}}}\bra{0_{\mathcal{X}}}$
and $\Delta_1 = I_{\mathcal{W}\otimes\mathcal{X}} - \Delta_0$.
The unitary operator
$I_{\mathcal{W}} \otimes 
(2 \ket{0_{\mathcal{X}}}\bra{0_\mathcal{X}} - I_{\mathcal{X}})$
performed during the simulation procedure may equivalently be written
$\Delta_0 - \Delta_1$.

At this point we will need to prove the claim that follows.
Although it is not difficult to prove, it is a key step in the analysis of
the simulator, and is isolated as a separate claim to highlight this fact.

\begin{claim} \label{claim:eigenvector}
The vector $\ket{\gamma_0}$ is an eigenvector
of the operator $\Delta_0 T^{\ast} V^{\ast} \Pi_0 V T \Delta_0$,
with corresponding eigenvalue $\lambda = 1/2$.
\end{claim}

\begin{proof}
The claim will hold regardless of the choice of $\ket{\psi}$.
To see that this is the case, define
\[
Q = (I_{\mathcal{W}} \otimes \bra{0_\mathcal{X}})T^{\ast} V^{\ast} \Pi_0 V T
(I_{\mathcal{W}} \otimes \ket{0_\mathcal{X}}).
\]
The operator $Q$ may be viewed as a measurement operator;
the pair $\{Q, I - Q\}$ describes the measurement that is effectively
performed on register $\reg{W}$ when the remaining registers are initialized,
the unitary transformation $VT$ is performed, and the measurement
$\{\Pi_0,\Pi_1\}$ is performed.
We have observed that this measurement outcome is 0 with probability 1/2,
so for every unit vector $\ket{\phi}\in\mathcal{W}$ we have
$\opbracket{\phi}{Q}{\phi} = 
\norm{\Pi_0 V T (\ket{\phi}\ket{0_{\mathcal{X}}})}^2 = \frac{1}{2}$.

Like every operator in $\lin{\mathcal{W}}$, the operator $Q$ is uniquely
determined by the function $\ket{\phi} \mapsto \opbracket{\phi}{Q}{\phi}$
defined on the unit sphere in $\mathcal{W}$, which implies that 
$Q = \frac{1}{2}I_{\mathcal{W}}$.
Therefore
\[
\Delta_0 T^{\ast} V^{\ast} \Pi_0 U T \Delta_0
=
(I_{\mathcal{W}}\otimes \ket{0_{\mathcal{X}}})
Q(I_{\mathcal{W}}\otimes \bra{0_{\mathcal{X}}})
=
\frac{1}{2}
I_{\mathcal{W}} \otimes \ket{0_{\mathcal{X}}}\bra{0_{\mathcal{W}}}.
\]
Clearly $\ket{\gamma_0} = \ket{\psi}\ket{0_{\mathcal{X}}}$ is an
eigenvector of this operator with corresponding eigenvalue $1/2$, which
completes the proof of the claim.
\end{proof}

Next, we will make use of a lemma that states a fact first proved
in \cite{MarriottW05}, where it was used to analyze an error reduction
method for the class $\class{QMA}$.
Because the proof is short it is included here for convenience.

\begin{lemma} \label{lemma:marriott-watrous}
Let $U,\Pi_0,\Pi_1,\Delta_0,\Delta_1\in\lin{\mathcal{H}}$ be linear operators
on a given Hilbert space $\mathcal{H}$ such that $U$ is unitary and $\Pi_0$,
$\Pi_1$, $\Delta_0$, and $\Delta_1$ are projections satisfying 
$\Delta_0 = I - \Delta_1$ and $\Pi_0 = I - \Pi_1$.
Suppose further that $\ket{\gamma_0}\in\mathcal{H}$ is a unit eigenvector of
$\Delta_0 U^{\ast} \Pi_0 U \Delta_0$ with corresponding eigenvalue
$\lambda \in (0,1)$.
Define
\[
\ket{\delta_0} = \frac{\Pi_0 U \ket{\gamma_0}}{\sqrt{\lambda}},\quad
\ket{\delta_1} = \frac{\Pi_1 U \ket{\gamma_0}}{\sqrt{1-\lambda}},
\quad\text{and}\quad
\ket{\gamma_1} = \frac{\Delta_1 U^{\ast} \ket{\delta_0}}{\sqrt{1-\lambda}}.
\]
Then
%$\ket{\gamma_1}$, $\ket{\delta_0}$ and $\ket{\delta_1}$ are unit vectors with
$\bracket{\gamma_0}{\gamma_1} = \bracket{\delta_0}{\delta_1} = 0$ and
\begin{align*}
U \ket{\gamma_0} & = 
\sqrt{\lambda} \ket{\delta_0} + \sqrt{1-\lambda} \ket{\delta_1},\\
U \ket{\gamma_1} & = 
\sqrt{1-\lambda} \ket{\delta_0} - \sqrt{\lambda} \ket{\delta_1}.
\end{align*}

\begin{proof}
First let us note that because $\ket{\gamma_0}$ is an eigenvector of 
$\Delta_0 U^{\ast} \Pi_0 U \Delta_0$ and the corresponding eigenvalue $\lambda$
is nonzero, it holds that $\Delta_0 \ket{\gamma_0} = \ket{\gamma_0}$.
By the definition of $\ket{\gamma_1}$, $\ket{\delta_0}$, and $\ket{\delta_1}$
it also holds that $\Delta_1 \ket{\gamma_1} = \ket{\gamma_1}$,
$\Pi_0 \ket{\delta_0} = \ket{\delta_0}$, and
$\Pi_1 \ket{\delta_1} = \ket{\delta_1}$.
Consequently $\bracket{\gamma_0}{\gamma_1} = \bracket{\delta_0}{\delta_1} = 0$.

The equation
$U \ket{\gamma_0} = \lambda \ket{\delta_0} + \sqrt{1 - \lambda} \ket{\delta_1}$
is immediate from the definitions of $\ket{\delta_0}$ and $\ket{\delta_1}$,
along with the fact that $\Pi_0=I-\Pi_1$.
Because
\[
\frac{\Delta_0 U^{\ast} \ket{\delta_0}}{\sqrt{\lambda}} =
\frac{\Delta_0 U^{\ast} \Pi_0 U \Delta_0 \ket{\gamma_0}}{\lambda} = 
\ket{\gamma_0},
\]
it also holds that
$U^{\ast} \ket{\delta_0}=\sqrt{\lambda}\ket{\gamma_0}
+ \sqrt{1-\lambda} \ket{\gamma_1}$, and thus
$U\ket{\gamma_1}=\sqrt{1-\lambda}\ket{\delta_0}-\sqrt{\lambda}\ket{\delta_1}$.
\end{proof}
\end{lemma}

It will be helpful when applying this lemma to note that for $U$ unitary
and $\lambda$ real, the following two sets of equations are equivalent:
\begin{xalignat*}{2}
U \ket{\gamma_0} & = 
\sqrt{\lambda} \ket{\delta_0} + \sqrt{1-\lambda} \ket{\delta_1}
&
U^{\ast} \ket{\delta_0} & = 
\sqrt{\lambda} \ket{\gamma_0} + \sqrt{1-\lambda} \ket{\gamma_1}\\
U \ket{\gamma_1} & = 
\sqrt{1-\lambda} \ket{\delta_0} - \sqrt{\lambda} \ket{\delta_1}
&
U^{\ast} \ket{\delta_1} & = 
\sqrt{1-\lambda} \ket{\gamma_0} - \sqrt{\lambda} \ket{\gamma_1}.
\end{xalignat*}

With Lemma~\ref{lemma:marriott-watrous} in hand, it now becomes simple to
analyze the behavior of the simulation procedure in the case where the
measurement outcome is 1.
Specifically, let us define
\[
\ket{\delta_0} = \sqrt{2} \Pi_0 V T \ket{\gamma_0},\quad
\ket{\delta_1} = \sqrt{2} \Pi_1 V T \ket{\gamma_0},
\quad\text{and}\quad
\ket{\gamma_1} = \sqrt{2} \Delta_1 T^{\ast} V^{\ast} \ket{\delta_0}.
\]
We have
$VT \ket{\gamma_0} = 
\frac{1}{\sqrt{2}} \ket{\delta_0} + \frac{1}{\sqrt{2}} \ket{\delta_1}$,
so that conditioned on the measurement outcome $0$ or $1$ the state of the
entire system becomes $\ket{\delta_0}$ or $\ket{\delta_1}$, respectively.
We have already observed that obtaining outcome 0 represents a successful
simulation, as
$\tr_{\mathcal{B}\otimes\mathcal{R}} \ket{\delta_0}\bra{\delta_0}
= \Phi(\ket{\psi}\bra{\psi})$
corresponds to the output of the actual interaction of $V$ with $P$.
In case the measurement outcome is 1 the state collapses to $\ket{\delta_1}$,
at which point the operators $(VT)^{\ast}$, $\Delta_0 - \Delta_1$, and
$VT$ are applied in sequence.
The operator $(VT)^{\ast}$ transforms the state $\ket{\delta_1}$ to
$\frac{1}{\sqrt{2}} \ket{\gamma_0} - \frac{1}{\sqrt{2}} \ket{\gamma_1}$,
the operator $\Delta_0 - \Delta_1$ transforms this state to
$\frac{1}{\sqrt{2}} \ket{\gamma_0} + \frac{1}{\sqrt{2}} \ket{\gamma_1}$,
and finally the operator $VT$ transforms this state to
$\frac{1}{2} \ket{\delta_0} +
\frac{1}{2} \ket{\delta_1} +
\frac{1}{2} \ket{\delta_0} -
\frac{1}{2} \ket{\delta_1} = \ket{\delta_0}$.
As in the case that the measurement outcome was 0, this state represents
a successful simulation.

We have thus established that the outcome of the simulation procedure is
precisely $\Phi(\ket{\psi}\bra{\psi})$ in case the initial state of $\reg{W}$
was $\ket{\psi}$.
Because the set
$\{\ket{\psi}\bra{\psi}\,:\,\ket{\psi}\in\mathcal{W},\,\norm{\ket{\psi}}=1\}$
spans all of $\lin{\mathcal{W}}$, and the map induced by the simulation
procedure is necessarily admissible (and therefore linear), it holds that
this map is precisely $\Phi$.
In other words, because admissible maps are uniquely determined by their
action on pure states, the map induced by the simulation procedure must be
$\Phi$; the simulation procedure implements {\it exactly} the same admissible
map as the actual interaction between $V$ and $P$.

%=============================================================================%

\section{Other Zero-Knowledge Proof Systems}
\label{sec:other}

The argument used in Section~\ref{sec:graph-isomorphism} to prove that the
Goldreich-Micali-Wigderson graph isomorphism protocol is zero-knowledge against
quantum attacks can be adapted to prove the same for several other protocols.
Some examples are discussed in this section---formal proofs concerning these
examples will appear in the final version of this paper.

%-----------------------------------------------------------------------------%

\subsection{Statistical zero-knowledge proof systems}

Let us begin with the simple observation that the proof in
Section~\ref{sec:graph-isomorphism} can easily be adapted to some other
protocols having a similar form to the protocol of Figure~\ref{fig:protocol},
meaning (i) $P$ sends a message to $V$, (ii) $V$ flips a fair coin and sends
the result to $P$, and (iii) $P$ responds with a second message.
An example of a protocol of this form that remains zero-knowledge under
quantum attacks is the quadratic residuosity protocol of
Goldwasser, Micali, and Rackoff \cite{GoldwasserM+89}.
The important aspects of such protocols that may allow the same proof to go
through with very little change is that in each case there exists a simulator
whose success probability is independent of the auxiliary input state of any
cheating quantum verifier.
This property translates into an analogous statement to
Claim~\ref{claim:eigenvector}, which then allows
Lemma~\ref{lemma:marriott-watrous} to be applied.

In the quantum setting, protocols of this simple form are universal for
honest-verifier quantum statistical zero-knowledge (with respect to the
definition given in \cite{Watrous02}), meaning that every problem having
a quantum interactive proof that is statistical zero-knowledge with respect to
an honest verifier also has a proof system of the above form.
Although such proof systems require the prover to send quantum information
to the verifier, and the verifier performs a quantum computation at the
end of the protocol, the verifier's single-bit message is classical.
(The honest prover can easily enforce this constraint just by measuring the
verifier's message before responding to it.)
This allows the proof from Section~\ref{sec:graph-isomorphism} to be easily
adapted to this setting as well.
Letting $\class{QSZK}_{\mathrm{HV}}$ denote the class of promise problems
having honest-verifier quantum statistical zero-knowledge protocols and 
$\class{QSZK}$ the class of problems that are quantum statistical
zero-knowledge with respect to the definitions we have discussed in
Section~\ref{sec:preliminaries}, we obtain the following corollary.

\begin{cor}
$\class{QSZK} = \class{QSZK}_{\mathrm{HV}}$.
\end{cor}

Although the statement of this corollary is analogous to the fact
$\class{SZK} = \class{SZK}_{\mathrm{HV}}$ of Goldreich, Sahai, and Vadhan
\cite{GoldreichS+98}, we hasten to add that the facts are only really similar
on the surface---there is no similarity in the proofs.
The quantum case is greatly simplified by the fact that every problem in
$\class{QSZK}_{\mathrm{HV}}$ has the very simple type of protocol discussed
above.
The class $\class{QSZK}_{\mathrm{HV}}$ does contain $\class{SZK}$, implying
that any problem having a classical statistical zero-knowledge proof system
(against classical verifiers) also has a quantum interactive proof system that
is statistical zero-knowledge against quantum verifiers.
Unfortunately, in the new proof system the honest prover and honest verifier
are required to perform quantum computations and the prover must send
quantum information to the verifier.
The question of whether every problem in $\class{SZK}$ has a classical
proof system that is zero-knowledge against quantum attacks is
not answered in this paper.

%-----------------------------------------------------------------------------%

\subsection{Computational zero-knowledge proof systems for $\class{NP}$}

Finally, we will discuss computational zero-knowledge protocols for any
problem in $\class{NP}$.
Specifically, we will consider the computational zero-knowledge proof for
Graph 3-Coloring (G3C) due to Goldreich, Micali, and Wigderson 
\cite{GoldreichM+91}.
Here, the input is a graph $G\in\mathcal{G}_n$, and the prover is attempting
to prove to the verifier that $G$ is 3-colorable.
A zero-knowledge proof system for this problem yields a zero-knowledge proof
for any problem in $\class{NP}$, as a protocol for an arbitrary $\class{NP}$
problem can begin with both parties computing a reduction to 3-coloring.
The fact that the zero-knowledge property is preserved under such a
reduction is discussed in \cite{GoldreichM+91}, and the quantum and classical
settings do not differ in this respect.
Specifically, the input to the original problem may be incorporated into
the verifier's auxiliary input, and therefore can offer no help in extracting
knowledge from the proof system's honest prover.

The protocol is based on the notion of a {\it commitment scheme}.
Because we will require a scheme that is meaningful in the presence of
quantum computation, it will be necessary for us to discuss various
issues concerning quantum computational indistinguishability and a formal
definition of the type of commitment scheme that will be required.

Before discussing these issues, it will be helpful to mention some conventions
and notation we will use regarding quantum circuits.
It will be convenient to allow quantum circuits to include two simple,
non-unitary gates: {\it ancillary} gates, which take no input and output a
single qubit in state $\ket{0}$, and {\it trace-out} gates that take one input
qubit and give no output, effectively throwing the qubit in the trash.
In addition to these two gates, quantum circuits may include Toffoli gates,
Hadamard gates, and imaginary-phase-shift gates (which induce the
transformation $\ket{0}\mapsto\ket{0}$ and $\ket{1}\mapsto i\ket{1}$),
which form a universal set of unitary gates.
A quantum circuit may therefore have a different number of input and output
qubits---we will say that a circuit is of type $(n,m)$ if it has $n$ input
qubits and $m$ output qubits.
More generally, an arbitrary admissible map from $n$ qubits to $m$ qubits will
be said to be of type $(n,m)$.
The {\it size} of a type $(n,m)$ quantum circuit is defined to be the number of
gates in the circuit plus $n+m$.
When $Q$ is such a circuit, we identify $Q$ with the admissible map from $n$
qubits to $m$ qubits induced by running $Q$.

%.............................................................................%

\subsubsection{Quantum computationally indistinguishability and zero-knowledge}

A quantum analogue of computational zero-knowledge requires a formal notion
of quantum computational indistinguishability.
Here we define one such notion, first for ensembles of states and then
for ensembles of admissible mappings.
In addition to forming the basis of our definition for quantum computational
zero-knowledge, the notion of quantum computational indistinguishability will
be required when we formalize the notion of a quantum computationally
concealing commitment scheme.

\begin{defn}\label{def:indistinguishable1}
{\bf (Polynomially quantum indistinguishable ensembles of states)}.
Let $S\subseteq\{0,1\}^{\ast}$ be an infinite set, let
$m:\{0,1\}^{\ast}\rightarrow\mathbb{N}$ be a polynomially bounded function,
and let $\rho_x$ and $\xi_x$ be mixed states on $m(x)$ qubits for each
$x\in S$.
Then the ensembles $\{\rho_x\,:\,x\in S\}$ and $\{\xi_x\,:\,x\in S\}$ are
{\it polynomially quantum indistinguishable} if, for every choice of
\begin{mylist}{\parindent}
\item[1.] polynomials $p$ and $q$,
\item[2.] a polynomially-bounded function 
$k:\{0,1\}^{\ast}\rightarrow\mathbb{N}$,
\item[3.] a collection $\{\sigma_x\,:\,x\in S\}$, where $\sigma_x$
is a mixed state on $k(x)$ qubits, and 
\item[4.] a quantum circuit $Q$ of size at most $p(|x|)$ and type
$(m(x) + k(x) , 1)$,
\end{mylist}
it holds that
\[
\abs{
  \opbracket{1}{Q(\rho_x\otimes\sigma_x)}{1} -
  \opbracket{1}{Q(\xi_x\otimes\sigma_x)}{1}} < \frac{1}{q(|x|)}
\]
for all but finitely many $x\in S$.
\end{defn}

When $\{\rho_n\,:\,n\in\mathbb{N}\}$ and $\{\xi_n\,:\,n\in\mathbb{N}\}$
are ensembles indexed by the natural numbers, we simply identify $S$
with $1^{\ast}$, interpreting each $n$ with its unary representation.
Let us also note that the above definition applies to the situation where
$\{\rho_x\,:\,x\in S\}$ and $\{\xi_x\,:\,x\in S\}$ represent classical
probability distributions, which are special cases of mixed states.

Notice that the above definition gives a fairly strict quantum analogue to the
typical non-uniform notion of classical polynomial indistinguishability.
It is strict because the non-uniformity includes an {\it arbitrary} quantum
state $\sigma_x$ that may aid some circuit $Q$ in the task of distinguishing
$\rho_x$ from~$\xi_x$.
In principle, this notion of non-uniformity is represented by the complexity
class $\class{BQP/qpoly}$ (see \cite{Aaronson05}).
In the present case, however, the ``advice'' state $\sigma_x$ may depend on
$x$ rather than just $|x|$, and we are interested in distinguishing
quantum states (or classical probability distributions) rather than
deciding language membership for strings.

An example where the state $\sigma_x$ plays an important role is as
follows.
Suppose $\rho_x$ and $\xi_x$ are {\it pure} and nearly orthogonal for
each $x$.
Then the ensembles $\{\rho_x\}$ and $\{\xi_x\}$ will fail to be polynomially
quantum indistinguishable, regardless of the complexity of the states;
taking $\sigma_x = \rho_x$, say, will allow a small circuit $Q$ to distinguish
$\rho_x$ and $\xi_x$ reasonably well by means of the ``swap test''
used in quantum fingerprinting \cite{BuhrmanC+01}.
The inclusion of the arbitrary state $\sigma_x$ is important in situations
(such as those we will consider in the context of zero-knowledge)
where indistinguishability of two ensembles must hold in the presence of other
``auxiliary'' information.

This definition is extended to admissible mappings by simply considering
ensembles that result from applying the mappings to arbitrary polynomial-size
states.

\begin{defn} {\bf (Polynomially quantum indistinguishable ensembles of
admissible maps)}.
Let $S\subseteq \{0,1\}^{\ast}$ be an infinite set and let
$\{\Phi_x\,:\,x\in S\}$
and $\{\Psi_x\,:\,x\in S\}$
be ensembles of admissible mappings indexed by $S$, where for each
$x\in S$ the mappings $\Phi_x$ and $\Psi_x$ are both of type $(n(x), m(x))$
for polynomially bounded functions $n$ and $m$.
Then these ensembles are {\it polynomially quantum indistinguishable}
if and only if, for every choice of
\begin{mylist}{\parindent}
\item[1.] polynomials $p$ and $q$,
\item[2.] a polynomially bounded function 
  $k: \{0,1\}^{\ast} \rightarrow \mathbb{N}$,
\item[3.] a collection of mixed states $\{\sigma_x\,:\,x\in S\}$, 
where $\sigma_x$ is a state on $n(x) + k(x)$ qubits, and 
\item[4.] a quantum circuit $Q$ of size at most $p(|x|)$ and type
  $(m(x) + k(x),1)$,
\end{mylist}
it holds that
\[
\abs{ \opbracket{1}{Q( (\Phi_x \otimes I)(\sigma_x))}{1}
- \opbracket{1}{Q( (\Psi_x \otimes I)(\sigma_x))}{1}}
< \frac{1}{q(|x|)}
\]
for all but finitely many $x\in S$.
\end{defn}

\noindent
(Note that a slight simplification is incorporated into this definition:
the input state $\sigma_x$ to the admissible mappings may include a part
that aids a given circuit $Q$ in distinguishing the outputs.)

Now we are prepared to state a definition for quantum computational
zero-knowledge.
Let $(V,P)$ be a proof system (quantum or classical) for a promise problem
$A = (A_{\mathrm{yes}},A_{\mathrm{no}})$.
This proof system will be said to be a 
{\it quantum computational zero-knowledge} for $A$ if, for every
polynomial-time quantum verifier $V'$ there exists a polynomial-time
quantum algorithm $S_{V'}$ that satisfies the following requirements.
Assume that on input $x$, the verifier $V'$ takes $n(x)$ auxiliary input
qubits and outputs $m(x)$ qubits, and let $\Phi_x$ denote the admissible
mapping of type $(n(x),m(x))$ that results from the interaction of $V'$
with $P$.
Then the simulator $S_{V'}$ must also take $n(x)$ qubits as input and output
$m(x)$ qubits, thereby implementing a mapping $\Psi_x$ of type
$(n(x),m(x))$.
Moreover, the ensembles $\left\{ \Phi_x\,:\,x\in A_{\mathrm{yes}}\right\}$
and $\left\{ \Psi_x\,:\,x\in A_{\mathrm{yes}}\right\}$ must be polynomially
quantum indistinguishable.

%.............................................................................%

\subsubsection{Quantum computationally concealing commitments}

The Goldreich-Micali-Wigderson G3C zero-knowledge proof system relies on the
prover's ability to commit to a 3-coloring of a given graph.
The binding property of these commitments is required for the soundness
of the proof system, while the concealing property is required for the
proof system to be zero-knowledge.
It is well-known that there cannot exist unconditionally binding and
concealing commitments based on quantum information alone \cite{Mayers97},
and therefore one must consider commitments for which either or both
of the binding and concealing properties is based on a computational
assumption.
In the interactive proof system setting, where one requires soundness
against arbitrary provers, the binding property of the commitments
must be unconditional, and therefore the concealing property must be
computationally-based.

Naturally, to be secure against quantum attacks, the commitment scheme
that is used must in fact be {\it quantum} computationally concealing.
The existence of such schemes is of course not proved, and does not follow
from the existence of classically computationally concealing commitment
schemes.
For example, good candidates for classically secure schemes based on the
computational difficulty of factoring or computing discrete logarithms
become insecure in the quantum setting because of Shor's
algorithm~\cite{Shor97}.
Classical commitments can, however, be based on arbitrary one-way functions
\cite{Naor91,HastadI+99}, and there are candidates for such functions that
may be difficult to invert even with efficient quantum algorithms.
Functions based on lattice problems, error-correcting codes, and non-abelian
group-theoretic problems represent candidates.

A protocol for quantum computationally concealing commitments based on the
existence of quantum one-way {\it permutations} is given in \cite{AdcockC02}.
Although the definitions in \cite{AdcockC02} differ somewhat from ours,
in particular in that they do not consider the stronger form of non-uniformity
allowing an auxiliary quantum state that we require, the result can be
translated to our setting.
(This naturally requires a somewhat stronger notion of a permutation being
one-way that forbids the possibility that a quantum circuit can invert
a one-way permutation using an auxiliary input.)

It should be noted that a protocol that is in some sense complementary to the
one in \cite{AdcockC02}, in that it is unconditionally concealing and
computationally binding, was given earlier in \cite{DumaisM+00}.
As mentioned above, this protocol seems not to be directly applicable to
the Goldreich-Micali-Wigderson G3C protocol because the binding property
can be broken by a computationally powerful party.
The protocol also requires quantum communication between honest parties,
although it may be possible to achieve the same result with only classical
communication and computation along the lines of \cite{NaorO+98}.

We now state our definition for the commitment schemes that we will require
to ensure the zero-knowledge property of the Goldreich-Micali-Wigderson
G3C protocol against quantum verifiers.
The definition is stated for an arbitrary finite set $\Gamma$;
the specific choice $\Gamma = \{1,2,3\}$ is used in the protocol.

\begin{defn}
Let $\Gamma$ be a finite set with $\abs{\Gamma} \geq 2$.
An {\it unconditionally binding, quantum computationally concealing
$\Gamma$-commitment scheme} consists of a deterministic polynomial-time
computable function $f$ with the following properties.
\begin{mylist}{\parindent}
\item[1.]
{\it (Uniform length.)}
There exists a polynomial $p$ such that
$\abs{f(a,x)} = p(\abs{x})$
for every $a\in \Gamma$ and $x\in\{0,1\}^{\ast}$.
(This requirement is not really essential, and is only made for convenience.)

\item[2.]
{\it (Binding property.)}
For every choice of $a\not=b\in\Gamma$ and $x,y\in\{0,1\}^{\ast}$,
we have $f(a,x) \not= f(b,y)$.

\item[3.]
{\it (Concealing property.)}
The ensembles $\{F_n(a)\,:\,n\in\mathbb{N}\}$ and 
$\{F_n(b)\,:\,n\in\mathbb{N}\}$ are polynomially quantum indistinguishable
for any choice of $a,b\in\Gamma$, where
$F_n(a)$ denotes the distribution obtained 
by evaluating $f(a,x)$ for $x\in\{0,1\}^n$ chosen uniformly at random.

\end{mylist}
\end{defn}

When such a scheme is used, it is assumed that some {\it security parameter}
$N$ is chosen---when one party (the prover in the G3C protocol) wishes to
commit to a value $a\in \Gamma$, a string $x\in \{0,1\}^N$ is chosen uniformly
at random and the string $f(a,x)$ is sent to the other party (the verifier
in the G3C protocol).
To reveal the commitment, the first party simply sends the string $x$ along
with the value $a$ to the second party, who checks the validity of the
decommitment by computing $f(a,x)$ and checking equality with the first
string sent.

%.............................................................................%

\subsubsection{The G3C protocol with perfect commitments}

Now we are ready to consider the zero-knowledge properties of the
Goldreich-Micali-Wigderson G3C protocol with respect to quantum verifiers.
It is helpful to begin by considering an idealized version of the protocol
assuming a {\it perfect} commitment scheme, meaning that the commitments
are unconditionally binding and concealing.
The proof system is described in Figure~\ref{fig:3-color}.

\begin{figure}[ht]
\noindent\hrulefill
\vspace{2mm}

{\bf\large\centerline{``Idealized'' Zero-Knowledge Protocol for G3C}}

\begin{trivlist}
\item Assume the input is a graph $G\in\mathcal{G}_n$ with $m$ edges.
  Repeat the following steps (sequentially) $m^2$ times:

\vspace{2mm}

\item {\bf Prover's step 1:}
  Let $\phi:\{1,\ldots,n\}\rightarrow \{1,2,3\}$ be a valid 3-coloring of $G$
  if one exists (otherwise let $\phi(u) = 1$ for each vertex $u$).
  Let $\pi\in S_3$ be a randomly generated permutation of the colors
  $\{1,2,3\}$.
  Commit to the values $\pi(\phi(1)),\ldots,\pi(\phi(n))$, sending these
  commitments to $V$.
  
\item {\bf Verifier's step 1:}
  Uniformly choose an edge $\{u,v\}$ of $G$ and send this edge to $P$.
  
\item {\bf Prover's step 2:}
  Reveal the values $\pi(\phi(u))$ and $\pi(\phi(v))$ to $V$.
  (Assume that every possible message from $V$ is decoded to a valid edge in
  $G$.)

\item {\bf Verifier's step 2:}
  Check that $\pi(\phi(u))\not=\pi(\phi(v))$, rejecting if not.

\vspace{2mm}

\item
  If the verifier has not rejected in any of the $m^2$ iterations, it
  accepts.

\end{trivlist}
\caption{The Goldreich-Micali-Wigderson G3C protocol,
  assuming perfect commitments.}
\label{fig:3-color}
\noindent\hrulefill
\end{figure}

A simulation procedure for this protocol for an arbitrary quantum 
polynomial-time verifier $V'$ can be constructed by simulating each
iteration of the loop individually.
Because each iteration allows an auxiliary quantum input, the zero-knowledge
property follows by a composition of the simulators.

One way to construct a {\it classical} simulator for each iteration of the
protocol is as follows.
The simulator uniformly chooses an edge $\{u,v\}$, and then selects
a function $\mu:\{1,\ldots,n\} \rightarrow \{1,2,3\}$ uniformly, subject to the
constraint that $\mu(u) \not= \mu(v)$.
The simulator computes commitments of the values $\mu(1),\ldots,\mu(n)$.
Although the function $\mu$ almost certainly does not constitute
a valid coloring of the graph $G$, the commitments of $\mu(1),\ldots,\mu(n)$
are indistinguishable from commitments of $\pi(\phi(1)),\ldots,\pi(\phi(n))$
for a valid coloring $\phi$.
The next step depends on the verifier $V'$ that the simulator is supposed to
simulate.
Given the commitments of $\mu(1),\ldots,\mu(n)$, along with whatever
auxiliary input it may have been given, the verifier $V'$ will choose some edge
$\{u',v'\}$.
In the idealized setting where the commitments ate {\it perfectly} concealing,
the choice of $\{u',v'\}$ agrees with $\{u,v\}$ with probability $1/m$.
Of course this will not necessarily be the case when the commitments are only
computationally concealing, which causes some technical complications that will
be addressed later.
In case $\{u,v\} = \{u',v'\}$, the commitments of $\mu(u)$ and $\mu(v)$ are
revealed, and the simulation of the current iteration is successful.
As for an actual interaction, the revealed colors are uniformly distributed
over the six possible distinct pairs of colors.
Otherwise, the entire process is repeated.
By repeating the process $O(m^2)$ times, say, the simulator is very likely to
obtain an iteration in which $\{u,v\} = \{u',v'\}$, representing a successful
simulation.
%Naturally, a formal argument (to be found in \cite{GoldreichM+91})
%is required to prove that the output resulting from the simulator is
%indistinguishable from the interaction between $V'$ and~$P$.

Now, based on such a classical simulation, we may define a quantum simulator
in a manner similar to the one in Section~\ref{sec:graph-isomorphism}.
We assume the verifier $V'$ has a similar set of registers to before,
except that now $\reg{A}$ stores an edge of $G$ rather than just a bit.
The unitary operator $T$ now represents a unitary implementation of the
first part of the classical simulation just described, with the register
$\reg{R}$ corresponding to all of the random bits that are needed for the
simulation.
(This will include the random choices used for the commitments when the
non-perfect commitments are discussed.)
The measurement $\{\Pi_0,\Pi_1\}$ now corresponds to testing that the registers
$\reg{A}$ and $\reg{B}$ contain the same edge.

In the ideal commitment case, Lemma~\ref{lemma:marriott-watrous} would be
applied with $\lambda = 1/m$.
Unlike the situation where $\lambda = 1/2$, however, as was the case for the
graph isomorphism protocol, we will now need to iterate a sequence of
measurements and unitary transformations in the main simulation procedure.
Instead of the simulation succeeding with certainty, this will allow
an exponentially small probability of failure.
The procedure is as follows:
\begin{mylist}{\parindent}
\item[1.]
Perform the unitary transformation $U = V T$.
\item[2.]
Perform the measurement $\{\Pi_0,\Pi_1\}$.
If the outcome is 0, the simulation has succeeded---halt and output
$(\reg{W},\reg{V},\reg{A},\reg{Y},\reg{Z})$.
\item[3.]
Perform the transformation $U^{\ast}$, followed by
$\Delta_0 - \Delta_1$, and then go to step 1.
\end{mylist}
This process may be terminated after some number of iterations depending
on the desired accuracy.

Supposing that states $\ket{\gamma_0}$, $\ket{\gamma_1}$, $\ket{\delta_0}$
and $\ket{\delta_1}$ are defined in an analogous way to the proof in 
Section~\ref{sec:graph-isomorphism}, the first application of $U$ maps the 
initial state $\ket{\gamma_0}$ to
\[
\frac{1}{\sqrt{m}}\ket{\delta_0}
+ \sqrt{\frac{m-1}{m}} \ket{\delta_1},
\]
and the measurement $\{\Pi_0,\Pi_1\}$ yields result 0 with probability
$1/m$.
Conditioned on result 0 the state becomes $\ket{\delta_0}$, which yields
a successful simulation.
Conditioned on result 1, which corresponds to the case where $V'$ has not
chosen the same edge as the simulator, the state becomes $\ket{\delta_1}$,
and the simulation continues.
The transformations $U^{\ast}$, $\Delta_0 - \Delta_1$, and $U$ are applied in
sequence, transforming the state $\ket{\delta_1}$ to
\[
\frac{2\sqrt{m-1}}{m} \ket{\delta_0} + \frac{m-2}{m}\ket{\delta_1}.
\]
The process continues in this way, with each iteration yielding a successful
simulation with probability at least $1/m$ (and somewhat more on each iteration
after the first).
After a polynomial number of iterations, the probability of a successful
simulation is therefore exponentially close to 1.

%.............................................................................%

\subsubsection{The G3C protocol with computational commitments}

Finally, we will briefly discuss the complications that arise when the
argument from the previous section is formalized when quantum computationally
concealing commitments replace the perfect commitments.
Our intention here is only to provide a sketch of the proof, highlighting
the aspects of the proof that differ from the classical case.
Further details will be included in the final version of the paper.

When a quantum computationally concealing commitment scheme is used in
place of the idealized, perfect commitment scheme in the protocol, there
must be a specified choice for the security parameter $N$.
It is sufficient to set $N$ to be equal to the number of vertices $n$
of the input graph for the purposes of establishing that the protocol is
quantum computational zero-knowledge.
The protocol is described in Figure~\ref{fig:3-color2}.

\begin{figure}[ht]
\noindent\hrulefill
\vspace{2mm}

{\bf\large\centerline{Computational Zero-Knowledge Protocol for 3-Coloring}}

\begin{trivlist}
\item Assume the input is a graph $G\in\mathcal{G}_n$ with $m$ edges.
  Also assume a quantum computationally concealing $\{1,2,3\}$-commitment
  scheme is given that is described by the function $f$.
  Repeat the following steps (sequentially) $m^2$ times:

\vspace{2mm}

\item {\bf Prover's step 1:}
  Let $\phi:\{1,\ldots,n\}\rightarrow \{1,2,3\}$ be a valid 3-coloring of $G$
  if one exists (otherwise let $\phi(u) = 1$ for each vertex $u$).
  Let $\pi\in S_3$ be a randomly generated permutation of the colors
  $\{1,2,3\}$.
  Choose strings $r_1,\ldots,r_n\in\{0,1\}^N$ uniformly at random and compute
  $s_u = f(\pi(\phi(u)),r_u)$ for each $u = 1,\ldots,n$.
  Send $s_1,\ldots,s_n$ to $V$.

\item {\bf Verifier's step 1:}
  Uniformly choose an edge $\{u,v\}$ of $G$ and send this edge to $P$.
  
\item {\bf Prover's step 2:}
  Send $V$ the values $a = \pi(\phi(u))$ and $b = \pi(\phi(v))$, along with the
  strings $r_u$ and $r_v$.
  (Assume that every possible message from $V$ is decoded to a valid edge in
  $G$.)

\item {\bf Verifier's step 2:}
  Check that $f(a,r_u) = s_u$, $f(b,r_v) = s_v$, and $a\not=b$,
  rejecting if not.

\vspace{2mm}

\item
  If the verifier has not rejected in any of the $m^2$ iterations, it
  accepts.

\end{trivlist}
\caption{The Goldreich-Micali-Wigderson G3C protocol, with computational
  commitments.}
\label{fig:3-color2}
\noindent\hrulefill
\end{figure}

Assume that a polynomial-time quantum verifier $V'$ for a single iteration of
the loop in the protocol is given.
As in the Graph Isomorphism case, it may be assumed without loss of generality
that $V'$ has a restricted form, being described by a collection of unitary
operators $\{V_y\,:\,y\in\{0,1\}^{n p(N)}\}$ acting on registers $\reg{W}$,
$\reg{V}$, $\reg{A}$, followed by a measurement of $\reg{A}$ in the standard
basis.
Here, $p(N)$ refers to the length of each commitment.
An arbitrary verifier for the entire protocol can be viewed as a composition
of such verifiers, possibly interleaved with polynomial-time quantum
computations---so a simulator for each such $V'$ allows for a simulation of
a general verifier.
The simulator for $V'$ will act precisely as described in the perfect
commitment case, substituting the computational commitments appropriately.

There are two main issues that arise in the proof.
The two issues are analogous to the issues arising in the classical
proof in \cite{GoldreichM+91}, but require some additional consideration
in the quantum setting.
The first issue concerns the running time of the simulator and the second
concerns the computational indistinguishability of the simulator's output
with the actual output of an interaction.

The first issue is the more difficult one.
The difficulty is that, because the commitments are not perfectly concealing,
it may no longer be assumed that every pure state $\ket{\psi}\in\mathcal{W}$ 
is an eigenvector of the operator
$Q = (I_{\mathcal{W}} \otimes \bra{0_{\mathcal{X}}})
T^{\ast} V^{\ast} \Pi_0 V T (I_{\mathcal{W}} \otimes \ket{0_{\mathcal{X}}})$.
This implies that Lemma~\ref{lemma:marriott-watrous} cannot be directly
applied.

The classical analogue to this issue is that the classical simulator may
have probability less than $1/m$ to ``guess'' the edge that will be asked by
a given $V'$.
Based on the fact that the commitments are computationally concealing,
one may argue that the probability of a correct ``guess'' deviates from
$1/m$ by some negligible quantity.
This involves a fairly straightforward {\it hybrid} argument: a significant
deviation in probability from $1/m$ in success could be turned into
an efficient procedure for breaking the concealing property of at least one of
the commitments.

In the quantum setting, a similar argument leads to the observation that
although the eigenvalues of the operator $Q$ might no longer all be equal 
to $1/m$, it must be the case that every eigenvalue of $Q$ is contained in
the interval
\[
\left(
\frac{1}{m} - \varepsilon(N),
\frac{1}{m} + \varepsilon(N)
\right)
\]
for $\varepsilon$ a negligible function.
An arbitrary auxiliary quantum input $\ket{\psi}$ may then be viewed as a
linear combination of eigenvectors of $Q$.
Each eigenvector gives rise to a different collection of vectors
$\ket{\delta_0}$, $\ket{\delta_1}$, $\ket{\gamma_0}$, and
$\ket{\gamma_1}$, which evolve independently but similarly as a result of the
simulation procedure.
The fact that the eigenvalues of $Q$ differ from one another by a negligible
quantity implies that this results in a negligible perturbation in the behavior
of the simulator from the perfect commitment case.

The second issue is more straightforward.
It must be demonstrated that the output of the simulator is computationally
indistinguishable from the output of an actual interaction (for an arbitrary
auxiliary input).
This can be handled by adapting the classical proof to the quantum setting.
Specifically, an efficient non-uniform procedure (i.e., a polynomial-size
quantum circuit together with an arbitrary auxiliary input) that distinguishes
the admissible maps corresponding to an actual interaction and the simulator
defined for a given verifier can be converted to a non-uniform procedure
that violates the concealing property of the commitment scheme, using
exactly the same type of hybrid argument as above.

%=============================================================================%

\section{Conclusion}
\label{sec:conclusion}

This paper has illustrated a method by which some interactive proof systems
can be proved to be zero-knowledge against quantum polynomial-time verifiers.
A few open questions and possible directions for further work in this
area are the following:

\begin{mylist}{\parindent}
\item[1.]
Although it has not been our aim to analyze as many classical zero-knowledge
protocols as possible using this technique, it may be beneficial to consider
further examples.
Possibly this will help to identify more general conditions under which
protocols can be proved zero-knowledge against quantum attacks.
A specific question along these lines is whether the statistical zero-knowledge
protocol that Goldreich, Sahai, and Vadhan \cite{GoldreichS+98} construct for
any given honest verifier statistical zero-knowledge proof system is 
zero-knowledge against quantum attacks.

\item[2.]
The variant of the Goldreich-Micali-Wigderson 3-coloring protocol that
is discussed in Section~\ref{sec:other} relies on the existence of quantum
computationally binding commitment schemes.
Such schemes follow from the existence of quantum one-way permutations
\cite{AdcockC02}.
The existence of such functions has the potential to become one of the most
important questions facing theoretical cryptography if quantum computers are
constructed.
What are the most promising candidates?

\item[3.]
There are several other variants of zero-knowledge such as concurrent and
resettable zero-knowledge.
Do the results of this paper have implications to quantum adversaries
in these settings?

\end{mylist}

%=============================================================================%

\subsection*{Acknowledgments}

I have had conversations and correspondences about quantum zero-knowledge with
several people, including Gilles Brassard, Richard Cleve, Claude Cr{\'e}peau, 
Simon-Pierre Desrosiers, Lance Fortnow, Dmitry Gavinsky, Dan Gottesman, 
Jordan Kerenidis, Hirotada Kobayashi, Ashwin Nayak, Amnon Ta-Shma, and
Alain Tapp, among others.
I thank all of them for their suggestions and input.
I would especially like to thank Claude Cr{\'e}peau for sharing his thoughts
and insight on zero-knowledge, and for getting me interested in the
problem discussed in this paper in the first place.
This research was supported by Canada's NSERC, the Canada Research Chairs
program, and the Canadian Institute for Advanced Research (CIAR).

%=============================================================================%

\bibliographystyle{alpha}
%\bibliography{Bibliography}

%=============================================================================%

\end{document}